\documentstyle[12pt,epsfig]{article}
                                                                                
\begin{document}                                                                
\date{}

\title{
{\vspace{-1em} \normalsize                                                      
\hfill \parbox{50mm}{DESY 99-036}}\\[25mm]
{\Large\bf Multi-bosonic algorithms for dynamical            \\
       fermion simulations}\footnote{Talk given at the workshop
       on {\em Molecular Dynamics on Parallel Computers},
       February 1999, NIC, J\"ulich; 
       to appear in the proceedings.}}

\author{I. Montvay                                           \\[0.5em]
        Deutsches Elektronen-Synchrotron DESY,               \\
        Notkestr.\,85, D-22603 Hamburg, Germany              \\
        E-mail: istvan.montvay@desy.de}

\newcommand{\be}{\begin{equation}}                                              
\newcommand{\ee}{\end{equation}}                                                
\newcommand{\half}{\frac{1}{2}}                                                 
\newcommand{\rar}{\rightarrow}                                                  
\newcommand{\lar}{\leftarrow}                                                   
                                                                                
\maketitle

\section{Introduction}\label{sec1}
 The numerical simulation of quantum field theories with fermions is an
 interesting and difficult computational problem.
 The basic difficulty is the necessary calculation of very large
 determinants, the determinants of {\em fermion matrices}, which can
 only be achieved by some stochastic procedure with the help of
 auxiliary bosonic ``pseudofermion'' fields.

 In hybrid Monte Carlo algorithms the number of pseudofermion fields
 corresponds to the number of fermion field components.
 The evolution of the pseudofermion fields in the updating process is
 realized by discretized molecular dynamics equations.
 The error implied by the finite length of discretization steps is
 corrected for by a global accept-reject decision which involves a
 fermion matrix inversion.
 The ingredients of the {\em two-step multi-bosonic
 algorithm}~\cite{TWO-STEP}, which will be considered in this
 contribution, are somewhat different but still in a general sense
 similar.
 The number of pseudofermion fields is multiplied by the order of a
 polynomial approximation of some negative power $x^{-\alpha}$ of the
 fermion matrix.
 These auxiliary bosonic fields are updated acording to a
 {\em multi-bosonic action}~\cite{LUSCHER} by using simple methods known
 from bosonic quantum field theories, as heatbath and overrelaxation.
 The error of the polynomial approximation is corrected also here in
 a global accept-reject decision.
 This is realized in a so called {\em noisy correction
 step}~\cite{NOISYCORR} by using a better polynomial approximation, which
 realizes a kind of generalized inversion of the fermion matrix.

 In my talk I review some recent developments of the multi-bosonic
 algorithms.
 The performance of the two-step multi-bosonic algorithm is illustrated
 in a recent large scale Monte Carlo simulation of the supersymmetric
 Yang-Mills theory~\cite{PHASETRANS,LIGHTGLUINO}, which is being performed
 at NIC, J\"ulich.
 This application shows that the algorithm is able to cope with the
 difficulties arising at nearly zero gluino masse in reasonably large
 physical volumes.

\section{Multi-bosonic actions}\label{sec2}

 The multi-bosonic representation of the fermion
 determinant~\cite{LUSCHER} is based on the approximation
\be\label{eq2_1}
\left|\det(Q)\right|^{N_f} =
\left\{\det(Q^\dagger Q) \right\}^{N_f/2}
\simeq \frac{1}{\det P_n(Q^\dagger Q)} \ ,
\ee
 where $N_f$ (Dirac-) fermion flavours are considered and the polynomial
 $P_n$ satisfies
\be\label{eq2_2}
\lim_{n \to \infty} P_n(x) = x^{-N_f/2}
\ee
 in an interval $x \in [\epsilon,\lambda]$.
 This interval is chosen such that it covers the spectrum of the squared
 hermitian fermion matrix $\tilde{Q}^2$.
 The hermitian matrix $\tilde{Q}$ is defined from the original fermion
 matrix $Q$
\be  \label{eq2_3}
\tilde{Q} \equiv \gamma_5 Q = \tilde{Q}^\dagger
\ee
 which is assumed here to satisfy the relation
\be  \label{eq2_4}
Q^\dagger = \gamma_5 Q \gamma_5 \ .
\ee
 Note that in (\ref{eq2_1}) only the absolute value of the determinant
 is taken which leaves out its sign (or phase).
 This can be taken into account at the evaluation of expectation values
 by reweighting.

 For the multi-bosonic representation of the determinant one uses
 the roots of the polynomial $r_j,\; (j=1,\ldots,n)$
\be\label{eq2_5}
P_n(Q^\dagger Q) = P_n(\tilde{Q}^2) = 
r_0 \prod_{j=1}^n (\tilde{Q}^2 - r_j) \ .
\ee
 Assuming that the roots occur in complex conjugate pairs, one can
 introduce the equivalent forms
\be\label{eq2_6}
P_n(\tilde{Q}^2) 
= r_0 \prod_{j=1}^n [(\tilde{Q} \pm \mu_j)^2 + \nu_j^2]
= r_0 \prod_{j=1}^n (\tilde{Q}-\rho_j^*) (\tilde{Q}-\rho_j)
\ee
 where $r_j \equiv (\mu_j+i\nu_j)^2$ and $\rho_j \equiv \mu_j + i\nu_j$.
 With the help of complex scalar (pseudofermion) fields $\Phi_{jx}$
 one can write
$$
\det P_n(Q^\dagger Q)^{-1} \propto
\prod_{j=1}^n\det[(\tilde{Q}-\rho_j^*) (\tilde{Q}-\rho_j)]^{-1}
$$
\vspace*{-1.2em}
\be\label{eq2_7}
\propto
\int [d\Phi]\; \exp\left\{ -\sum_{j=1}^n \sum_{xy}
\Phi_{jy}^+\, [(\tilde{Q}-\rho_j^*) (\tilde{Q}-\rho_j)]_{yx}\, 
\Phi_{jx} \right\} \ .
\ee
 The exponent in (\ref{eq2_7}) is the (negative) multi-bosonic action.
 Since it is quadratic in $\tilde{Q}$, its locality properties are
 inherited from the fermion matrix $Q$.
 For instance, if $Q$ has only nearest neighbour interactions then
 the multi-bosonic action (\ref{eq2_7}) extends up to next-to-nearest
 neighbours. 

 The multi-bosonic representation of the fermion determinant
 (\ref{eq2_7}) can be used for a Monte Carlo procedure in terms of the
 pseudofermion fields $\Phi_{jx}$.
 The difficulty for small fermion masses in large physical volumes is
 that the {\em condition number} $\lambda/\epsilon$ becomes very large
 ($10^4-10^6$) and very high orders $n = {\cal O}(10^3)$ are needed for
 a good approximation.
 This requires large storage and the autocorrelation becomes bad since
 it is proportional to $n$.
 An additional question is how to control the systematic errors
 introduced by the polynomial approximation in (\ref{eq2_1}).
 In principle one has to perform the limit to infinite order of the
 approximation $n\to\infty$, which means in practical terms a need to
 investigate the dependence of the results on $n$.

 Several solutions for eliminating the systematical errors due to the
 finite order of approximation $n$ in (\ref{eq2_1}) are possible.
 For instance, one can calculate a correction factor from the
 eigenvalues of $\tilde{Q}^2$ and introduce a corresponding global
 Matropolis accept-reject step in the updating~\cite{EIGENVALUE}.
 The necessary calculation of the eigenvalues leads, however, to an
 algorithm growing with the square of the number of lattice points.

 A better solution is to apply a noisy correction step~\cite{NOISYCORR}
 which is especially simple in case of $N_f=2$ flavours when an
 iterative inversion is sufficient~\cite{UCORRECTION}.
 This can be generalized to an arbitrary number of flavours in the
 two-step multi-bosonic scheme~\cite{TWO-STEP} which will be discussed
 in detail in the next section.
 The special case of $N_f=1$ flavours can be dealt with in a
 non-hermitian version~\cite{UCORRECTION,NON-HERMITIAN} applied directly
 to $Q$, insted of $Q^\dagger Q$ in (\ref{eq2_1}).
 This works well for heavy fermion masses~\cite{ONE-FLAVOUR} when the
 spectrum of $Q$ can be covered by en ellipse but it would be very
 cumbersome for small fermion masses where the eigenvalues are
 surrounding zero.

 Another possibility to perform the corrections of the systematic errors
 of the polynomial approximation is reweighting in the expectation
 values.
 For the special case of $N_f=2$ this has been introduced in the
 polynomial hybrid Monte Carlo scheme~\cite{MCORRECTION}.
 The case of arbitrary $N_f$ can be solved by an appropriate polynomial
 approximation, which can also be implemented in the two-step
 multi-bosonic approach (see section~\ref{sec3.2}).

 Other approaches for eliminating the systematic errors are possible,
 for instance, by choosing some specific ways of optimizing the
 approximate polynomials (see~\cite{WOLFF,FORCRAND}).
 It is also possible to combine the multi-bosonic idea with other
 methods of dynamical fermion simulations.
 To review all attempts would take too much time and most of the
 proposals were not yet tested in really large scale simulations.
 In what follows I shall concentrate on the two-step multi-bosonic
 scheme which proved to be efficient in recent simulations of SU(2)
 Yang-Mills theory with light gluinos~\cite{PHASETRANS,LIGHTGLUINO}.

\section{Two-step multi-bosonic algorithm}\label{sec3}

 The dynamical fermion algorithm directly using the multi-bosonic
 representation in (\ref{eq2_7}) has difficulties with large storage
 requirements and long autocorrelations.
 One can achieve substantial improvements on both these problems by
 introducing a two-step polynomial
 approximation~\cite{TWO-STEP,POLYNOM}.
 In this two-step approximation scheme (\ref{eq2_2}) is replaced by
\be\label{eq3_1}
\lim_{n_2 \to \infty} P^{(1)}_{n_1}(x)P^{(2)}_{n_2}(x) = 
x^{-N_f/2} \ , \hspace{3em} 
x \in [\epsilon,\lambda] \ .
\ee
 The multi-bosonic representation is only used for the first
 polynomial $P^{(1)}_{n_1}$ wich provides a first crude approximation
 and hence the order $n_1$ can remain relatively low.
 The correction factor $P^{(2)}_{n_2}$ is realized in a stochastic
 {\em noisy correction step} with a global accept-reject decision
 during the updating process (see section \ref{sec3.1}).
 In order to obtain an exact algorithm one has to consider in this case
 the limit $n_2\to\infty$.
 For very small fermion masses it turned out more practicable to fix
 some large $n_2$ and perform another small correction in the evaluation
 of expectation values by {\em reweighting} with a still finer
 polynomial (see section \ref{sec3.2}).

\subsection{Update correction: global accept-reject}\label{sec3.1}
 The idea to use a stochastic correction step in the
 updating~\cite{NOISYCORR}, instead of taking very large polynomial
 orders $n$, was proposed in the case of $N_f=2$ flavours
 in~\cite{UCORRECTION}.
 $N_f=2$ is special because the function to be approximated is just
 $x^{-1}$ and $P^{(2)}_{n_2}(x)$ can be replaced by the calculation of
 the inverse of $xP^{(1)}_{n_1}(x)$.
 For general $N_f$ one can take the two-step approximation scheme
 introduced in~\cite{TWO-STEP} where the two-step multi-bosonic
 algorithm is described in detail.
 The theory of the necessary optimized polynomials is summarized in
 section \ref{sec4} following~\cite{POLYNOM}.

 In the two-step approximation scheme for $N_f$ flavours of fermions
 the absolute value of the determinant is represented as
\be\label{eq3.1_1}
\left|\det(Q)\right|^{N_f} \;\simeq\;
\frac{1}{\det P^{(1)}_{n_1}(\tilde{Q}^2)  
\det P^{(2)}_{n_2}(\tilde{Q}^2)} \ .
\ee
 The multi-bosonic updating with $n_1$ scalar pseudofermion fields
 is performed by heatbath and overrelaxation sweeps for the scalar
 fields and Metropolis sweeps for the gauge field.
 After a Metropolis sweep for the gauge field a global accept-reject
 step is introduced in order to reach the distribution of gauge field
 variables $[U]$ corresponding to the right hand side of
 (\ref{eq3.1_1}).
 The idea of the noisy correction is to generate a random vector
 $\eta$ according to the normalized Gaussian distribution
\be \label{eq3.1_2}
\frac{e^{-\eta^\dagger P^{(2)}_{n_2}(\tilde{Q}[U]^2)\eta}}
{\int [d\eta] e^{-\eta^\dagger P^{(2)}_{n_2}(\tilde{Q}[U]^2)\eta}}  \ ,
\ee
 and to accept the change $[U^\prime] \lar [U]$ with probability
\be \label{eq3.1_3}
\min\left\{ 1,A(\eta;[U^\prime] \lar [U]) \right\} \ ,
\ee
 where
\be \label{eq3.1_4}
A(\eta;[U^\prime] \lar [U]) =
\exp\left\{-\eta^\dagger P^{(2)}_{n_2}(\tilde{Q}[U^\prime]^2)\eta -
            \eta^\dagger P^{(2)}_{n_2}(\tilde{Q}[U]^2)\eta\right\}\ .
\ee

 The Gaussian noise vector $\eta$ can be obtained from $\eta^\prime$
 distributed according to the simple Gaussian distribution
\be \label{eq3.1_5}
\frac{e^{-\eta^{\prime\dagger}\eta^\prime}}
{\int [d\eta^\prime] e^{-\eta^{\prime\dagger}\eta^\prime}}
\ee
 by setting it equal to
\be \label{eq3.1_6}
\eta = P^{(2)}_{n_2}(\tilde{Q}[U]^2)^{-\half} \eta^\prime  \ .
\ee

 In order to obtain the inverse square root on the right hand side of
 (\ref{eq3.1_6}), we can proceed with polynomial approximations in two
 different ways.
 The first possibility was proposed in~\cite{TWO-STEP} with 
 $x \equiv \tilde{Q}^2$ as
\be \label{eq3.1_7}
P^{(2)}_{n_2}(x)^{-\half} \simeq R_{n_3}(x) \simeq x^{N_f/4}
S_{n_s}[P^{(1)}_{n_1}(x)] \ .
\ee
 Here 
\be \label{eq3.1_8}
S_{n_s}(P) \simeq  P^\half
\ee
 is an approximation of the function $P^\half$ on the interval
 $P \in [\lambda^{-N_f/2},\epsilon^{-N_f/2}]$.
 The polynomial approximations $R_{n_3}$ and $S_{n_s}$ can be determined
 by the same general procedure as $P^{(1)}_{n_1}$ and $P^{(2)}_{n_2}$.
 It turns out that these approximations are ``easier'' in the sense that
 for a given order higher precisions can be achieved than, say, for
 $P^{(1)}_{n_1}$.

 Another possibility to obtain a suitable approximation for 
 (\ref{eq3.1_6}) is to use the second decomposition in (\ref{eq2_6})
 and define
\be\label{eq3.1_9}
P^{(1/2)}_{n_2}(\tilde{Q}) \equiv 
\sqrt{r_0} \prod_{j=1}^{n_2}(\tilde{Q}-\rho_j) \ ,\hspace{3em}
P^{(2)}_{n_2}(\tilde{Q}^2) = 
P^{(1/2)}_{n_2}(\tilde{Q})^\dagger P^{(1/2)}_{n_2}(\tilde{Q}) \ .
\ee
 Using this form, the noise vector $\eta$ necessary in the noisy
 correction step can be generated from the gaussian vector
 $\eta^\prime$ according to
\be\label{eq3.1_10}
\eta=P^{(1/2)}_{n_2}(\tilde{Q})^{-1}\eta^\prime \ ,
\ee
 where $P^{(1/2)}_{n_2}(\tilde{Q})^{-1}$ can be obtained as
\be\label{eq3.1_11}
P^{(1/2)}_{n_2}(\tilde{Q})^{-1} = 
\frac{P^{(1/2)}_{n_2}(\tilde{Q})^\dagger}
{P^{(2)}_{n_2}(\tilde{Q}^2)} \simeq 
P_{n_3}(\tilde{Q}^2) P^{(1/2)}_{n_2}(\tilde{Q})^\dagger \ .
\ee
 In the last step $P_{n_3}$ denotes a polynomial approximation for
 the inverse of $P^{(2)}_{n_2}$ on the interval $[\epsilon,\lambda]$.
 Note that this last approximation can also be replaced by an iterative
 inversion of $P^{(2)}_{n_2}(\tilde{Q}^2)$.
 However, tests showed that the inversion by a least-squares optimized
 polynomial approximation is much faster because, for a given precision,
 less matrix multiplications have to be performed.

 In the simulation with light dynamical
 gluinos~\cite{PHASETRANS,LIGHTGLUINO} mainly the second form in
 (\ref{eq3.1_10})-(\ref{eq3.1_11}) has been used.
 The first form could, however, be used as well.
 In fact, for very high orders $n_2$ or on a 32-bit computer the first
 scheme is better from the point of view of rounding errors.
 The reason is that in the second scheme for the evaluation of
 $P^{(1/2)}_{n_2}(\tilde{Q})$ we have to use the product form in terms
 of the roots $\rho_j$ in (\ref{eq3.1_9}).
 Even using the optimized ordering of roots defined
 in~\cite{TWO-STEP,POLYNOM}, this is numerically less stable than the
 recursive evaluation according to (\ref{eq4.1_4}), (\ref{eq4.1_10}).
 If one uses the first scheme both $P^{(2)}_{n_2}$ in (\ref{eq3.1_4})
 and $R_{n_3}$ in (\ref{eq3.1_6})-(\ref{eq3.1_7}) can be evaluated
 recursively.
 Nevertheless, on a 64-bit machine both methods work well and in case
 of (\ref{eq3.1_11}) the determination of the least-squares optimized
 polynomials is somewhat simpler.

 The global accept-reject step for the gauge field can be performed
 after full sweeps over the gauge field links.
 A good choice for the order $n_1$ of the first polynomial
 $P^{(1)}_{n_1}$ is such that the average acceptance probability of the
 noisy correction be near 90\%.
 One can decrease $n_1$ and/or increase the acceptance probability by
 updating only some subsets of the links before the accept-reject step.
 This might be useful on lattices larger than the largest lattice
 $12^3 \cdot 24$ considered in~\cite{LIGHTGLUINO}.

\subsection{Measurement correction: reweighting}\label{sec3.2}
 The multi-bosonic algorithms become exact only in the limit of
 infinitely high polynomial orders: $n\to\infty$ in (\ref{eq2_2}) or, in
 the two-step approximation scheme,  $n_2\to\infty$ in (\ref{eq3_1}).
 Instead of investigating the dependence on the polynomial order by
 performing several simulations, it is practically better to fix some
 high order for the simulation and perform another correction in the
 ``measurement'' of expectation values by still finer polynomials.
 This is done by {\em reweighting} the configurations in the measurement
 of different quantities.
 In case of $N_f=2$ flavours this kind of reweighting has been used
 in~\cite{MCORRECTION} within the polynomial hybrid Monte Carlo scheme.
 As remarked above, $N_f=2$ is special because the reweighting can be
 performed by an iterative inversion.
 The general case can, however, also be treated by a further polynomial
 approximation.

 The measurement correction for general $N_f$ has been introduced
 in~\cite{BOULDER}.
 It is based on a polynomial approximation $P^{(4)}_{n_4}$ which
 satisfies
\be\label{eq3.2_1}
\lim_{n_4 \to \infty} P^{(1)}_{n_1}(x)P^{(2)}_{n_2}(x)P^{(4)}_{n_4}(x) =
x^{-N_f/2} \ , \hspace{3em} 
x \in [\epsilon^\prime,\lambda] \ .
\ee
 The interval $[\epsilon^\prime,\lambda]$ can be chosen, for instance,
 such that $\epsilon^\prime=0,\lambda=\lambda_{max}$, where
 $\lambda_{max}$ is an absolute upper bound of the eigenvalues of
 $Q^\dagger Q=\tilde{Q}^2$.
 In this case the limit $n_4\to\infty$ is exact on an arbitrary gauge
 configuration.
 For the evaluation of $P^{(4)}_{n_4}$ one can use $n_4$-independent
 recursive relations (see section \ref{sec4}), which can be stopped
 by observing the convergence of the result.
 After reweighting the expectation value of a quantity $A$ is given by
\be\label{eq3.2_2}
\langle A \rangle = \frac{
\langle A \exp{\{\eta^\dagger[1-P^{(4)}_{n_4}(Q^\dagger Q)]\eta\}}
\rangle_{U,\eta}}
{\langle  \exp{\{\eta^\dagger[1-P^{(4)}_{n_4}(Q^\dagger Q)]\eta\}}
\rangle_{U,\eta}} \ ,
\ee
 where $\eta$ is a simple Gaussian noise like $\eta^\prime$ in
 (\ref{eq3.1_5}).
 Here $\langle\ldots\rangle_{U,\eta}$ denotes an expectation value
 on the gauge field sequence, which is obtained in the two-step process
 described in the previous subsection, and on a sequence of independent
 $\eta$'s.
 The expectation value with respect to the $\eta$-sequence can be
 considered as a Monte Carlo updating process with the trivial action
 $S_\eta \equiv \eta^\dagger\eta$.
 The length of the $\eta$-sequence on a fixed gauge configuration can
 be, in principle, arbitrarily chosen.
 In praxis it can be optimized for obtaining the smallest possible
 errors.

 The application of the measurement correction is most important for
 quantities which are sensitive for small eigenvalues of the fermion
 matrix $Q^\dagger Q$.
 The polynomial approximations are worst near $x=0$ where the function
 $x^{-N_f/2}$ diverges.
 In the exact effective gauge action, including the fermion determinant,
 the configuration with a small eigenvalue $\Lambda$ are suppressed by
 $\Lambda^{N_f/2}$.
 The polynomials at finite order are not able to provide such a strong
 suppression, therefore in the updating sequence of the gauge fields
 there are more configurations with small eigenvalues than needed.
 The {\em exceptional configurations} with exceptionally small
 eigenvalues have to be supressed by the reweighting.
 This can be achieved by choosing $\epsilon^\prime=0$ and a high enough
 order $n_4$.
 It is also possible to take some non-zero $\epsilon^\prime$ and
 determine the eigenvalues below it exactly.
 Each eigenvalue $\Lambda < \epsilon^\prime$ is taken into account by an
 additional reweighting factor 
 $\Lambda^{N_f/2}P^{(1)}_{n_1}(\Lambda)P^{(2)}_{n_2}(\Lambda)$.
 The stochastic correction in (\ref{eq3.2_2}) is then restricted to the
 subspace orthogonal to these eigenvectors.
 Instead of $\epsilon^\prime > 0$ one can also keep $\epsilon^\prime=0$
 and project out a fixed number of smallest eigenvalues.

 Let us note that, in principle, it would be enough to perform just a
 single kind of correction.
 But to omit the reweighting does not pay because it is much more
 comfortable to investigate the (small) effects of different $n_4$
 values on the expectation values than to perform several simulations
 with increasing values of $n_2$.
 Without the updating correction the whole correction could be done by
 reweighting in the measurements.
 However, in practice this would not work either.
 The reason is that a first polynomial with relatively low order does
 not sufficiently suppress the exceptional configurations.
 As a consequence, the reweighting factors would become too small and
 would reduce the effective statistics considerably.
 In addition, the very small eigenvalues are changing slowly in the
 update and this would imply longer autocorrelations.

 A moderate surplus of gauge configurations with small eigenvalues may,
 however, be advantageous because it allows for an easier tunneling
 among sectors with different topological charges.
 For small fermion masses on large physical volumes this is expected to
 be more important than the prize one has to pay for it by reweighting,
 provided that the reweighting has only a moderate effect.

\section{Least-squares optimized polynomials}\label{sec4}

 The basic ingredient of multi-bosonic fermion algorithms is the
 construction of the necessary optimized polynomial approximations.
 The least-squares optimization provides a general and flexible
 framework which is well suited for the requirements of multi-bosonic
 algorithms~\cite{POLYNOM}.
 In the first part of this section the necessary basic formulae are
 collected.
 In the second part a simple example is considered:  in case of an
 appropriately chosen weight function the least-squares optimized
 polynomials for the approximation of the function $x^{-\alpha}$ are
 expressed in terms of Jacobi polynomials.

\subsection{Definition and basic relations}\label{sec4.1}
 The general theory of least-squares optimized polynomial approximations
 can be inferred from the literature~\cite{FOXPARKER,RIVLIN}.
 Here we introduce the basic formulae in the way it has been done
 in~\cite{POLYNOM} for the specific needs of multi-bosonic fermion
 algorithms.
 We shall keep the notations there, apart from a few changes which
 allow for more generality.

 We want to approximate the real function $f(x)$ in the interval
 $x \in [\epsilon,\lambda]$ by a polynomial $P_n(x)$ of degree $n$.
 The aim is to minimize the deviation norm
\be \label{eq4.1_1}
\delta_n \equiv \left\{ N_{\epsilon,\lambda}^{-1}\;
\int_\epsilon^\lambda dx\, w(x)^2 \left[ f(x) - P_n(x) \right]^2
\right\}^\half \ .
\ee
 Here $w(x)$ is an arbitrary real weight function and the overall
 normalization factor $N_{\epsilon,\lambda}$ can be chosen by
 convenience, for instance, as
\be \label{eq4.1_2}
N_{\epsilon,\lambda} \equiv
\int_\epsilon^\lambda dx\, w(x)^2 f(x)^2 \ .
\ee
 A typical example of functions to be approximated is
 $f(x)=x^{-\alpha}/\bar{P}(x)$ with $\alpha > 0$ and some polynomial
 $\bar{P}(x)$.
 The interval is usually such that $0 \leq \epsilon < \lambda$.
 For optimizing the relative deviation one takes a weight function
 $w(x) = f(x)^{-1}$.

 It turns out useful to introduce orthogonal polynomials
 $\Phi_\mu(x)\; (\mu=0,1,2,\ldots)$ satisfying
\be \label{eq4.1_3}
\int_\epsilon^\lambda dx\, w(x)^2 \Phi_\mu(x)\Phi_\nu(x) 
= \delta_{\mu\nu} q_\nu \ .
\ee
 and expand the polynomial $P_n(x)$ in terms of them:
\be \label{eq4.1_4}
P_n(x) = \sum_{\nu=0}^n d_{n\nu} \Phi_\nu(x) \ .
\ee
 Besides the normalization factor $q_\nu$ let us also introduce, for
 later purposes, the integrals $p_\nu$ and $s_\nu$ by
\be \label{eq4.1_5}
q_\nu \equiv \int_\epsilon^\lambda dx\, w(x)^2 \Phi_\nu(x)^2 \ ,
\hspace{1em}
p_\nu \equiv \int_\epsilon^\lambda dx\, w(x)^2 \Phi_\nu(x)^2 x \ ,
\hspace{1em}
s_\nu \equiv \int_\epsilon^\lambda dx\, w(x)^2 x^\nu \ .
\ee

 It can be easily shown that the expansion coefficients $d_{n\nu}$ 
 minimizing $\delta_n$ are independent from $n$ and are given by
\be \label{eq4.1_6}
d_{n\nu} \equiv d_\nu = \frac{b_\nu}{q_\nu} \ ,
\ee
 where
\be \label{eq4.1_7}
b_\nu \equiv \int_\epsilon^\lambda dx\, w(x)^2 f(x) \Phi_\nu(x) \ .
\ee
 The minimal value of $\delta_n^2$ is
\be \label{eq4.1_8}
\delta_n^2 = 1 - N_{\epsilon,\lambda}^{-1} 
\sum_{\nu=0}^n d_\nu b_\nu \ .
\ee

 The above orthogonal polynomials satisfy three-term recurrence
 relations which are very useful for numerical evaluation.
 The first two of them with $\mu=0,1$ are given by
\be \label{eq4.1_9}
\Phi_0(x) = 1 \ , \hspace{2em}
\Phi_1(x) = x - \frac{s_1}{s_0} \ .
\ee
 The higher order polynomials $\Phi_\mu(x)$ for $\mu=2,3,\ldots$ can be
 obtained from the recurrence relation
\be \label{eq4.1_10}
\Phi_{\mu+1}(x) = (x+\beta_\mu)\Phi_\mu(x) + 
\gamma_{\mu-1}\Phi_{\mu-1}(x) \ ,
\hspace{2em} (\mu=1,2,\ldots) \ ,
\ee
 where the recurrence coefficients are given by
\be \label{eq4.1_11}
\beta_\mu = -\frac{p_\mu}{q_\mu} \ , \hspace{3em}
\gamma_{\mu-1} = -\frac{q_\mu}{q_{\mu-1}} \ .
\ee

 Using these relations on can set up a recursive scheme for the
 computation of the orthogonal polynomials in terms of the basic
 integrals $s_\nu$ defined in (\ref{eq4.1_5}).
 Defining the polynomial coefficients $f_{\mu\nu}\; (0\leq\nu\leq\mu)$
 by
\be \label{eq4.1_12}
\Phi_\mu(x) = \sum_{\nu=0}^\mu f_{\mu\nu}x^{\mu-\nu} 
\ee
 the above recurrence relations imply the normalization convention
\be \label{eq4.1_13}
f_{\mu 0} = 1 \ , \hspace{3em} (\mu=0,1,2,\ldots) \ ,
\ee
 and one can easily show that $q_\mu$ and $p_\mu$ satisfy
\be \label{eq4.1_14}
q_\mu = \sum_{\nu=0}^\mu f_{\mu\nu} s_{2\mu-\nu} \ ,
\hspace{2em}
p_\mu = \sum_{\nu=0}^\mu f_{\mu\nu}
\left( s_{2\mu+1-\nu} +f_{\mu 1}s_{2\mu-\nu} \right) \ .
\ee
 The coefficients themselves can be calculated from $f_{11}=-s_1/s_0$
 and (\ref{eq4.1_10}) which gives
\begin{eqnarray}\label{eq4.1_15}
f_{\mu+1,1} & = & f_{\mu,1} + \beta_\mu \ ,             \nonumber \\
f_{\mu+1,2} & = & f_{\mu,2} + \beta_\mu f_{\mu,1} + 
\gamma_{\mu-1} \ ,                                      \nonumber \\
f_{\mu+1,3} & = & f_{\mu,3} + \beta_\mu f_{\mu,2} + 
\gamma_{\mu-1} f_{\mu-1,1} \ ,                          \nonumber \\
            & \ldots &                                  \nonumber \\
f_{\mu+1,\mu} & = & f_{\mu,\mu} + \beta_\mu f_{\mu,\mu-1} + 
\gamma_{\mu-1} f_{\mu-1,\mu-2} \ ,                      \nonumber \\
f_{\mu+1,\mu+1} & = & \beta_\mu f_{\mu,\mu} + 
\gamma_{\mu-1} f_{\mu-1,\mu-1} \ .
\end{eqnarray}
 The polynomial and recurrence coefficients are recursively determined
 by (\ref{eq4.1_13})-(\ref{eq4.1_15}).
 The expansion coefficients for the optimized polynomial $P_n(x)$
 can be obtained from (\ref{eq4.1_6}) and
\be \label{eq4.1_16}
b_\mu = \sum_{\nu=0}^\mu f_{\mu\nu}
\int_\epsilon^\lambda dx\, w(x)^2 f(x) x^{\mu-\nu} \ .
\ee
%

\subsection{A simple example: Jacobi polynomials}\label{sec4.2}
 The approximation interval $[\epsilon,\lambda]$ can be transformed to
 some standard interval, say, $[-1,1]$ by the linear mapping
\be \label{eq4.2_1}
\xi = \frac{2x-\lambda-\epsilon}{\lambda-\epsilon} \ , \hspace{2em}
x = \frac{\xi}{2}(\lambda-\epsilon) + \half(\lambda+\epsilon) \ .
\ee
 A weight factor $(1+\xi)^\rho(1-\xi)^\sigma$ with $\rho,\sigma > -1$
 corresponds in the original interval to the weight factor
\be \label{eq4.2_2}
w^{(\rho,\sigma)}(x)^2 = (x-\epsilon)^\rho (\lambda-x)^\sigma \ . 
\ee
 Taking, for instance, $\rho=2\alpha,\; \sigma=0$ this weight is similar
 to the one for relative deviation from the function $f(x)=x^{-\alpha}$,
 which would be just $x^{2\alpha}$.
 In fact, for $\epsilon=0$ these are exactly the same and for small
 $\epsilon$ the difference is negligible.
 The advantage of considering the weight factor in (\ref{eq4.2_2}) is that
 the corresponding orthogonal polynomials are simply related to the
 Jacobi polynomials~\cite{FREUD,RIGR}, namely
\be \label{eq4.2_3}
\Phi^{(\rho,\sigma)}_\nu(x) = (\lambda-\epsilon)^\nu \nu!\,
\frac{\Gamma(\rho+\sigma+\nu+1)}{\Gamma(\rho+\sigma+2\nu+1)}
P^{(\sigma,\rho)}_\nu
\left(\frac{2x-\lambda-\epsilon}{\lambda-\epsilon}\right) \ .
\ee
 Our normalization convention (\ref{eq4.1_13}) implies that
\be \label{eq4.2_4}
q^{(\rho,\sigma)}_\nu = (\lambda-\epsilon)^{\rho+\sigma+2\nu+1}\nu!\,
\frac{\Gamma(\rho+\nu+1)\Gamma(\sigma+\nu+1)\Gamma(\rho+\sigma+\nu+1)}
{\Gamma(\rho+\sigma+2\nu+1)\Gamma(\rho+\sigma+2\nu+2)} \ .
\ee
 The coefficients of the orthogonal polynomials are now given by
\be \label{eq4.2_5}
f^{(\rho,\sigma)}_{\mu\nu} = \sum_{\omega=0}^\nu 
(-\epsilon)^{\nu-\omega}(\epsilon-\lambda)^\omega
\left(\begin{array}{c} \mu-\omega \\ \nu-\omega \end{array}\right)
\left(\begin{array}{c} \mu \\ \omega \end{array}\right)
\frac{\Gamma(\rho+\mu+1)\Gamma(\rho+\sigma+2\mu-\omega+1)}
{\Gamma(\rho+\mu-\omega+1)\Gamma(\rho+\sigma+2\mu+1)} \ . 
\ee
 In particular, we have
\be \label{eq4.2_6}
f^{(\rho,\sigma)}_{\mu0} = 1 \ , \hspace{2em}
f^{(\rho,\sigma)}_{11} = -\epsilon - (\lambda-\epsilon)
\frac{(\rho+1)}{(\rho+\sigma+2)} \ .
\ee
 The coefficients $\beta,\gamma$ in the recurrence relation (\ref{eq4.1_10})
 can be derived from the known recurrence relations of the Jacobi
 polynomials:
$$
\beta^{(\rho,\sigma)}_\mu = -\half(\lambda+\epsilon) +
\frac{(\sigma^2-\rho^2)(\lambda-\epsilon)}
{2(\rho+\sigma+2\mu)(\rho+\sigma+2\mu+2)} \ ,
$$
\be \label{eq4.2_7}
\gamma^{(\rho,\sigma)}_{\mu-1} = -(\lambda-\epsilon)^2\,
\frac{\mu(\rho+\mu)(\sigma+\mu)(\rho+\sigma+\mu)}
{(\rho+\sigma+2\mu-1)(\rho+\sigma+2\mu)^2(\rho+\sigma+2\mu+1)} \ .
\ee

 In order to obtain the expansion coefficients of the least-squares 
 optimized polynomials one has to perform the integrals in (\ref{eq4.1_16}).
 As an example, let us consider the function $f(x)=x^{-\alpha}$ when
 the necessery integrals can be expressed by hypergeometric functions:
$$
\int_\epsilon^\lambda dx\, (x-\epsilon)^\rho (\lambda-x)^\sigma
x^{\mu-\nu-\alpha} =
$$
\be \label{eq4.2_8}
= (\lambda-\epsilon)^{\rho+\sigma+1} \lambda^{\mu-\nu-\alpha}
\frac{\Gamma(\rho+1)\Gamma(\sigma+1)}{\Gamma(\rho+\sigma+2)}
F\left( \alpha-\mu+\nu,\sigma+1;\rho+\sigma+2;
1-\frac{\epsilon}{\lambda} \right) \ .
\ee
 Let us now consider, for simplicity, only the case $\epsilon=0$, when
 we obtain
\be \label{eq4.2_9}
b^{(\rho,\sigma)}_\mu = (-1)^\mu \lambda^{1+\rho+\sigma+\mu-\alpha}
\frac{\Gamma(\rho+\sigma+\mu+1)\Gamma(\alpha+\mu)\Gamma(\rho-\alpha+1)
\Gamma(\sigma+\mu+1)}{\Gamma(\rho+\sigma+2\mu+1)\Gamma(\alpha)
\Gamma(\rho+\sigma-\alpha+\mu+2)} \ .
\ee
 Combined with (\ref{eq4.1_6}) and (\ref{eq4.2_4}) this leads to
\be \label{eq4.2_10}
d^{(\rho,\sigma)}_\mu = (-1)^\mu \lambda^{-\mu-\alpha}
\frac{\Gamma(\rho+\sigma+2\mu+2)\Gamma(\alpha+\mu)\Gamma(\rho-\alpha+1)}
{\mu!\,\Gamma(\rho+\mu+1)\Gamma(\alpha)\Gamma(\rho+\sigma-\alpha+\mu+2)}
 \ .
\ee

 These formulae can be used, for instance, for fractional inversion.
 For the parameters $\rho,\sigma$ the natural choice in this case is
 $\rho=2\alpha,\sigma=0$ which corresponds to the optimization of the
 relative deviation from the function $f(x)=x^{-\alpha}$.
 As we have seen in section \ref{sec4.1}, the optimized polynomials
 are the truncated expansions of $x^{-\alpha}$ in terms of the Jacobi
 polynomials $P^{(2\alpha,0)}$.
 The Gegenbauer polynomials proposed in~\cite{BUNK} for fractional
 inversion correspond to a different choice, namely
 $\rho=\sigma=\alpha-\half$.
 This is because of the relation
\be \label{eq4.2_11}
C^\alpha_n(x) = \frac{\Gamma(n+2\alpha)\Gamma(\alpha+\half)}
{\Gamma(2\alpha)\Gamma(n+\alpha+\half)} 
P^{(\alpha-\half,\alpha-\half)}_n(x) \ .
\ee
 Note that for the simple case $\alpha=1$ we have here the Chebyshev
 polynomials of second kind: $C^1_n(x)=U_n(x)$.

 The special case $\alpha=\half$ is interesting for the numerical
 evaluation of the zero mass lattice action proposed by
 Neuberger~\cite{NEUBERGER}.
 In this case, in order to obtain the least-squares optimized relative
 deviation with weight function $w(x)=x$, the function $x^{-\half}$ has
 to be expanded in the Jacobi polynomials $P^{(1,0)}$.
 Note that this is different both from the Chebyshev and the Legendre
 expansions applied in~\cite{HEJALU}.
 The former would correspond to take $P^{(-\half,-\half)}$, the latter
 to $P^{(0,0)}$.
 The corresponding weight functions would be $[x(\lambda-x)]^{-\half}$
 and $1$, respectively.
 As a consequence of the divergence of the weight factor at $x=0$,
 the Chebyshev expansion is not appropriate for an approximation in an
 interval with $\epsilon=0$.
 This can be immediately seen from the divergence of
 $d^{(-\half,-\half)}_\mu$ at $\alpha=\half$ in (\ref{eq4.2_10}).

 The advantage of the Jacobi polynomials appearing in these examples is
 that they are analytically known.
 The more general least-squares optimized polynomials defined in the
 previous subsection can also be numerically expanded in terms of them.
 This is sometimes more comfortable than the entirely numerical
 approach.

\section{Outlook}\label{sec5}
 
 The two-step multi-bosonic algorithm has been shown to work properly
 in a recent large scale numerical simulation of light dynamical
 gluinos~\cite{PHASETRANS,LIGHTGLUINO}.
 Since gluinos are Majorana fermions the effective number of (Dirac-)
 fermion flavours is in this case $N_f=\half$.
 This simulation is demanding because it deals with nearly zero and
 negative fermion masses in reasonably large physical volumes.
 It turned out possible to investigate the expected first order phase
 transition at zero gluino mass.
 The algorithm was able to cope with the metastability of phases near
 the phase transition.
 Up to now the investigated physical volumes were not very large:
 in case of the largest ($12^3 \cdot 24$) lattice the product of the
 spatial extension times the square root of the string tension was
 $L \sqrt{\sigma} \simeq 2.4$.
 The experience shows, however, that simulations with light gluinos and
 $L \sqrt{\sigma} \simeq 5$ or larger would be feasible with reasonable
 effort.
 Another important conclusion in~\cite{LIGHTGLUINO} is that the
 inclusion of the sign of the fermion determinant, actually Pfaffian for
 Majorana fermions, can be achieved by determining the spectral flow of
 the hermitian fermion matrix $\tilde{Q}$ as a function of increasing
 hopping parameter.

 An interesting application of the two-step multi-bosonic algorithm is
 the numerical simulation of QCD.
 Up to now no serious effort was taken in this direction but first
 tests will be performed soon.
 A particularly relevant aspect for QCD is the possibility to deal with
 an odd number of quark flavours ($N_f$).
 The popular hybrid Monte Carlo algorithm can only be applied for even
 $N_f$.
 For non-even $N_f$ one can use finite step-size molucular dynamics
 algorithms like the one in \cite{MOLECULAR}, but then the extrapolation
 to zero step size is an additional difficulty.



\begin{thebibliography}{99}
%
\bibitem{TWO-STEP}
I. Montvay,
{\it Nucl. Phys.} B {\bf 466,} 259 (1996).
%
\bibitem{LUSCHER}
M. L\"uscher,
{\it Nucl. Phys.} B {\bf 418,} 637 (1994).
%
\bibitem{NOISYCORR}
A.D. Kennedy, J. Kuti,
{\it Phys. Rev. Lett.} {\bf 54,} 2473 (1985); \\
A.D. Kennedy, J. Kuti, S. Meyer, B.J. Pendleton,
{\it Phys. Rev.} D {\bf 38,} 627 (1988).
%
\bibitem{PHASETRANS}
R. Kirchner, S. Luckmann, I. Montvay, K. Spanderen, J. Westphalen,
{\it Phys. Lett.} B {\bf 446,} 209 (1999).
%
\bibitem{LIGHTGLUINO}
I. Campos, A. Feo, R. Kirchner, S. Luckmann, I. Montvay, G. M\"unster,
K. Spanderen, J. Westphalen,
hep-lat/9903014.
%
\bibitem{EIGENVALUE}
C. Alexandrou, A. Borrelli, P. de Forcrand, A. Galli, F. Jegerlehner,
{\it Nucl. Phys.} B {\bf 456,} 296 (1995).
%
\bibitem{UCORRECTION}
A. Bori\c{c}i, P. de Forcrand,
{\it Nucl. Phys.} B {\bf 454,} 645 (1995).
%
\bibitem{NON-HERMITIAN}
A. Borrelli, P. de Forcrand and A. Galli,
{\it Nucl. Phys.} B {\bf 477,} 809 (1996).
%
\bibitem{ONE-FLAVOUR}
C. Alexandrou, A. Borici, A. Feo, P. de Forcrand, A. Galli, F. Jegerlehner,
T. Takaishi,
{\it Nucl. Phys. Proc. Suppl.} B {\bf 53,} 435 (1997);
{\it Nucl. Phys. Proc. Suppl.} B {\bf 63,} 406 (1998); 
hep-lat/9811028.
%
\bibitem{MCORRECTION}
R. Frezzotti, K. Jansen,
{\it Phys. Lett.} B {\bf 402,} 328 (1997);
hep-lat/9808011; hep-lat/9808038.
%
\bibitem{WOLFF}
U. Wolff,
{\it Nucl. Phys. Proc. Suppl.} B {\bf 63,} 937 (1998).
%
\bibitem{FORCRAND}
P. de Forcrand,
hep-lat/9809145. 
%
\bibitem{POLYNOM}
I. Montvay,
{\it Comput. Phys. Commun.} {\bf 109,} 144 (1998).
%
\bibitem{BOULDER}
R. Kirchner, S. Luckmann, I. Montvay, K. Spanderen, J. Westphalen,
to appear in the proceedings of the Lattice '98 Conference, Boulder,
July 1998, hep-lat/9808024.
%
\bibitem{SPANDEREN}
K. Spanderen,
{\em Monte Carlo simulations of SU(2) Yang-Mills theory with dynamical
gluinos,} PhD Thesis, University M\"unster, August 1998 (in German).
%
\bibitem{FOXPARKER}
L. Fox, I.B. Parker,
{\em Chebyshev Polynomials in Numerical Analysis,} Oxford University
Press, 1968.
%
\bibitem{RIVLIN}
T.J. Rivlin,
{\em An Introduction to the Approximation of Functions,}
Blaisdell Publ. Company, 1969.
%
\bibitem{FREUD}
G. Freud,
{\em Orthogonale Polynome,}
Birkh\"auser Verlag, 1969.
%
\bibitem{RIGR}
I.S. Gradshteyn, I.M. Ryzhik,
{\em Table of Integrals, Series, and Products,}
Academic Press, 1965.
%
\bibitem{BUNK}
B. Bunk,
{\it Nucl. Phys. Proc. Suppl.} B {\bf 63,} 952 (1998).
%
\bibitem{NEUBERGER}
H. Neuberger,
{\it Phys. Lett.} B {\bf 417} 141 (1998).
%
\bibitem{HEJALU}
P. Hernandez, K. Jansen, M. L\"uscher,
hep-lat/9808010. 
%
\bibitem{MOLECULAR}
S. Gottlieb, W. Liu, D. Toussaint, R.L. Renken, R.L. Sugar,
{\it Phys. Rev.} D {\bf 35,} 2531 (1987).
%
\end{thebibliography}
\end{document}